\documentstyle{article}

\def\draft#1{} \def\preprint#1{\small#1} \def\submit#1{} 

\def\be{\begin{equation}} \def\ee{\end{equation}}

\def\arcsinh{\mathop{\rm arcsinh}\nolimits}

\renewcommand{\d}{{\rm d}}
\newcommand{\case}[2]{{\textstyle\frac{#1}{#2}}}
\newcommand{\ms}{\noalign{\vspace{3pt plus2pt minus1pt}}}
\newcommand{\bs}{\noalign{\vspace{6pt plus2pt minus2pt}}}

\def\threeV{\,{}^{(3)}\negthinspace V}

\def\Nscr{{\cal N}} \def\Pscr{{\cal P}} \def\Qscr{{\cal Q}}
\newcommand{\scri}{\Im}

\def\undersim#1{\mathop{\vtop{\ialign{##\crcr
     $\hfil\displaystyle{#1}\hfil$\crcr\noalign
     {\kern1pt\nointerlineskip}\hbox{$\hfil\sim\hfil$}\crcr
     \noalign{\kern1pt}}}}}

\catcode`@=11

\def\eqalign#1{\null\,\vcenter{\openup\jot\m@th
  \ialign{\strut\hfil$\displaystyle{##}$&$\displaystyle{{}##}$\hfil
      \crcr#1\crcr}}\,}
\def\meqalign#1{\null\,\vcenter{\openup\jot\m@th
  \ialign{\strut\hfil$\displaystyle{##}$&&$\displaystyle{{}##}$\hfil
      \crcr#1\crcr}}\,}
\catcode`@=12   

\def\umlaut{\"} 

\def\journalfont{\rm}         
\def\jou#1{{\journalfont #1\ }}
\def\joudef#1#2{\def #1{\jou{\ignorespaces #2}}}

\joudef{\aaa}    { Astron.\ Astrophys.}
\joudef{\aip}    { Adv.\ Phys.}
\joudef{\adm}    { Adv.\ Math.}
\joudef{\am}     { Ann.\ Math.}
\joudef{\apny}   { Ann.\ Phys.\ (N.Y.)}
\joudef{\apj}    { Astrophys.\ J.}
\joudef{\cjp}    { Can.\ J.\ Phys.}
\joudef{\cmp}    { Commun.\ Math.\ Phys.}
\joudef{\cqg}    { Class.\ Quantum Grav.}
\joudef{\grg}    { Gen.\ Rel.\ Grav.}
\joudef{\ijmpd}  { Int.\ J.\ Mod.\ Phys.\ D}
\joudef{\ijtp}   { Int.\ J.\ Theor.\ Phys.}
\joudef{\im}     { Invent.\ Math.}
\joudef{\jm}     { J.\ Math.}
\joudef{\jmp}    { J.\ Math.\ Phys.}
\joudef{\jpa}    { J.\ Phys.\ A}
\joudef{\mnras}  { Mon.\ Not.\ R.\ Ast.\ Soc.}
\joudef{\mpla}   { Mod.\ Phys.\ Lett.\ A}
\joudef{\nature} { Nature}
\joudef{\nc}     { Nuovo Cim.}
\joudef{\npb}    { Nuc.\ Phys.\ B}
\joudef{\pla}    { Phys.\ Lett. A}
\joudef{\plb}    { Phys.\ Lett. B}
\joudef{\pr}     { Phys.\ Rev.}
\joudef{\prd}    { Phys.\ Rev.\ D}
\joudef{\prep}   { Phys.\ Rep.}
\joudef{\prl}    { Phys.\ Rev.\ Lett.}
\joudef{\prsla}  { Proc.\ Roy.\ Soc.\ Lond.\ A}
\joudef{\ptp}    { Prog.\ Theor.\ Phys.}
\joudef\rmp      { Rev.\ Mod.\ Phys.}
\joudef\spj      { Sov.\ Phys.\ JETP}

\begin{document}
\bibliographystyle{prsty}

\preprint{Submitted to Classical and Quantum Gravity
         \hfill Preprint, gr-qc/9411047, November 1994}

\vfill
\centerline{\bf\Large Hamiltonian approach to relativistic star models}
\vskip1truecm
\centerline{\large Kjell Rosquist\footnote{E-mail: kr@physto.se}}
\medskip
\centerline{\sl Department of Physics, Stockholm University,
                Box 6730, 113 85 Stockholm, Sweden}
\centerline{and}
\centerline{\sl RGGR group, Chimie-Physique, CP231, Universit\'e Libre
            de Bruxelles,}
\centerline{\sl Campus Plaine CP 231, 1050 Brussels, Belgium}
\vfill

\begin{abstract}
An ADM-like Hamiltonian approach is proposed for static spherically symmetric
relativistic star configurations. For a given equation of state the entire
information about the model can be encoded in a certain 2-dimensional
minisuperspace geometry. We derive exact solutions which arise from
symmetries corresponding to linear and quadratic geodesic invariants in
minisuperspace by exploiting the relation to minisuperspace Killing tensors.
A classification of exact solutions having the full number of integrations
constants is given according to their minisuperspace symmetry properties. In
particular it is shown that Schwarzschild's exterior solution and Buchdahl's
$n=1$ polytrope solution correspond to minisuperspaces with a Killing vector
symmetry while Schwarzschild's interior solution, Whittaker's solution and
Buchdahl's $n=5$ polytrope solution correspond to minisuperspaces with a
second rank Killing tensor. New solutions filling in empty slots in this
classification scheme are also given. One of these new solutions has a
physically reasonable equation of state and is a generalization of Buchdahl's
$n=1$ polytrope model.
\end{abstract}

\vfill\clearpage
\section{Introduction}
The subject of relativistic star configurations goes back to the very birth of
general relativity itself when Schwarzschild gave his interior solution in
1916 \cite{schwarz:interior}. Although a large body of literature now exists
on the subject there still remain unsolved problems. For example, a
theoretical understanding of some aspects of the interplay between the
equation of state and the geometry of the model has only recently been
achieved \cite{lm:indexlimit,simon:polyfive}. The stability of stellar models
is another area in which only partial results are known concerning the
relation between stability and the equation of state. In this paper we present
a Hamiltonian formulation of the equations for spherically symmetric static
equilibrium configurations. The Hamiltonian framework is then applied to show
how known exact solutions are explained in terms of Killing tensor symmetries
in a certain minisuperspace. A number of new exact solutions are also
uncovered in this process.

Spherically symmetric gravitational fields belong to a class of models in
which the spacetime is foliated or sliced by a family of homogeneous
hypersurfaces. The causal character of the slicing is timelike. These facts
lead one to compare and contrast these models with another class of models,
namely the spatially homogeneous models, also hypersurface homogeneous but
with a spacelike slicing. From a mathematical point of view the difference
between star models and spatially homogeneous spacetimes is that the
independent variable is a spacelike radial coordinate for stars while for
spatially homogeneous models it is a timelike coordinate which is interpreted
as a cosmic time function. The dramatic success and usefulness of Hamiltonian
methods for exact solutions \cite{urj:geom,ru:kt}, solution structure
\cite{ujrz:late}, quantum cosmology \cite{atu:miniquant} and visualization of
the minisuperspace dynamics \cite{ru:visual} make it natural to ask if the
Hamiltonian methods might also be useful for star models. In fact the ADM
formalism works also for timelike slicings of the spacetime
\cite{mtw:gravitation}. In place of the lapse function one must then
introduce a radial gauge function. One consequence of this is that the
Schwarzschild coordinates are unsuitable as a starting point for the
Hamiltonian formulation. Instead they will correspond to a certain choice of
momentum dependent radial gauge function. Also, as explained below one must
be careful when introducing the matter terms to obtain a correct Hamiltonian.
This paper is intended to serve two main purposes the first of which is to
outline the Hamiltonian approach to star models in general while the second
main purpose is to classify those exact solutions which are related to
symmetries of the relativistic equations for static equilibrium.

The importance of the Hamiltonian approach for exact solutions stems from
the possibility to define a dynamical minisuperspace geometry which carries
all information about the dynamics. We shall define such a dynamical
geometry by using a special radial gauge function to be referred to as the
Jacobi gauge. This is analogous to the Jacobi lapse which was introduced
for spatially homogenous models in \cite{ruj:vac,ruj:mat}. The resulting
Jacobi geometry is conformally related to minisuperspace geometries
corresponding to other choices of lapse function. Given such a dynamical
geometry the problem of finding and classifying exact solutions becomes a
matter of studying geodesic orbits, a subject which has been treated from
many points of view in the literature. There are two different mechanisms by
which exact solutions can occur. A system of differential equations can be
reduced either by means of a symmetry or by means of an invariant submanifold
leading to symmetry solutions and submanifold solutions, corresponding to the
classical notions of general and particular solutions respectively. Broadly
speaking, symmetry solutions are more useful since they allow one to impose
arbitrary initial conditions. Fortunately, they are also more amenable to
systematic study by using the theory of Lie symmetries for differential
equations.

In the context of relativistic star models initial conditions correspond to
matching conditions at the surface of the star and possibly inside the star
to match the solution to one with a different equation of state above a
certain pressure. For spherically symmetric models the matching condition is
simply that the pressures of the two solutions must be equal at the boundary.
In particular, the surface of the star is defined as the point of zero
pressure and this automatically matches the solution to the Schwarzschild
vacuum solution by Birkhoff's theorem \cite{mtw:gravitation}. In addition, a
solution which goes all the way to the center of the star must be regular
there. This is the well-known condition of elementary flatness. It is
discussed in section \ref{sec:ham} in the context of the Hamiltonian
framework. An alternative to using symmetry solutions is to assume a certain
functional form for the radial dependence of a metric component (e.g.\
\cite{tolman:statsol}) or for the energy density (e.g.\
\cite{knutsen:gasspheres}). If the functional form involves some arbitrary
parameters, then one may obtain physically reasonable submanifold solutions
by adjusting those parameters in order to satisfy the boundary conditions.

In the geometric formulation of the dynamics%
\footnote{The word dynamics is  somewhat inappropriate here since we
are dealing with static models. However, the radial variable will be seen to
play a role which is identical from the mathematical point of view to the
role played by the time in the closely related spatially homogeneous models.
In particular the radial metric coordinate serves as the independent variable
for the Hamiltonian. Lacking a more suitable term we will refer to the
Hamiltonian equations of star configurations as ``dynamics".}
the Hamiltonian takes the purely kinetic form $H_J = \frac12 J^{AB} p_A p_B$
where the Jacobi metric $J_{AB}$ is the metric of minisuperspace in the
Jacobi gauge. In this picture the star configurations correspond to the
timelike geodesics of the Jacobi geometry. The simplest symmetries of $H_J$
are those which are related to Killing vectors of
$J_{AB}$. It is well-known that a Killing vector $\xi^A$ gives
rise to a linear constant of the motion for $H_J$ given by $\xi^A
p_A$. Similarly, constants of the motion may exist which are
homogeneous polynomials of quadratic or higher degree in the momenta. In the
quadratic case such constants of the motion have the form $\xi^{AB}
p_A p_B$ where $\xi^{AB}$ are the components of a
geometric object known as a second rank Killing tensor \cite{ksmh:exact}. In
general, a constant of the motion of the $n$'th degree corresponds to a $n$'th
rank Killing tensor. In particular a Killing vector is a Killing
tensor of rank one. The precise relation between Killing tensors and
Lie symmetries of the geodesic equations was given in \cite{rosquist:kt_sym}.

In the ADM formulation the static star configurations are 1+1-dimensional
Hamiltonian systems. A criterion for the existence of Killing vectors and
second rank Killing tensors for such systems was given in \cite{ru:kt}. It
was also shown in \cite{ru:kt} how a Killing vector or a Killing tensor can
be used to obtain variables which are adapted to the symmetry. Such symmetry
adapted variables are needed for explicit integration of the equations of
motion. There is a large number of exact solutions in the literature (see
\cite{ksmh:exact} for a partial list) some of them symmetry solutions and
some submanifold solutions. One of our main objectives is to identify the
symmetry solutions.

We shall use the Killing tensor criterion of \cite{ru:kt}
to look for integrable relativistic star models. While a general solution to
this problem is not available a present, even for Killing tensors up to
second rank, we will be able to make substantial progress by using a certain
ansatz for the symmetry adapted variables. The resulting family of solutions
includes all (to this author's knowledge) known symmetry solutions as well
as some new solutions. There is no guarantee that the symmetry solutions are
physically reasonable. We shall therefore use the following rough criterion to
separate solutions with a more reasonable equation of state from the
obviously unphysical solutions. We consider the equation of state to
be physical if the conditions $p \geq 0$, $\rho \geq 0$ and $0 \leq \d
p/ \d\rho \leq 1$ are satisfied. The last of these conditions ensures that
the velocity of sound is well-defined by $v_{\rm sound} = \sqrt{\d p/ \d\rho}$
and that it does not exceed the speed of light. Note that an equation of state
which is defined by some function $p = f(\rho)$ may be physical at some
densities but unphysical at others. Our condition is in fact rather strict
and excludes for example the incompressible fluid used in Schwarzschild's
interior solution. One of the new solutions found in this work is
physical according to our criterion and generalizes Buchdahl's
$n=1$ polytrope solution \cite{buchdahl:poly_one}.

\section{Hamiltonian formulation of static models}
\label{sec:ham}

We consider static spherically symmetric matter configurations where the
matter is a perfect fluid described by some equation of state $p =
f(\rho)$. The stress-energy tensor is given by $T_{\alpha\beta} =
(\rho+p)u_\alpha u_\beta + p g_{\alpha\beta}$ and the metric is required to
satisfy the Einstein equations $G_{\alpha\beta} = \kappa T_{\alpha\beta}$.
The metric of spherically symmetric models is usually written as
\cite{mtw:gravitation}
\be\label{eq:schwarzvar}
     ds^2 = -e^{2\nu} dt^2 + e^{2\lambda} dr^2 + r^2 \d\Omega^2 \ ,
\ee
where
\be
     \d\Omega^2 = \d\theta^2 + \sin^2 \theta \,\d\phi^2 \ ,
\ee
is the metric of the 2-sphere. One reason why this form has been preferred in
the past is that the Schwarzschild $r$ variable is invariantly defined by its
relation $A=4\pi r^2$ to the area $A$ of a 2-surface given by $dt=dr=0$. One
lesson to be learned, though, from the experience of spatially homogeneous
models is that no preferred set of variables exists which is suitable for all
situations. In particular, for the Hamiltonian method we need to modify the
form (\ref{eq:schwarzvar}) of the metric. Using the analogy with spatially
homogeneous models a natural starting point for a Hamiltonian formulation is
to write the metric in the form
\be\label{eq:metric}
        ds^2 = -e^{2\beta^3} dt^2 + N^2 dR^2
              + e^{2\beta^1} \d\Omega^2 \ ,
\ee
where $N$ is a radial gauge function analogous to the lapse function in
formulations with a spacelike slicing. For static models $N$ is a function
of the radial coordinate $R$. The relation to conventional variables is given
be the relations
\be
     \beta^3 = \nu\ ,\qquad e^{\beta^1} = r\ , \qquad NdR = e^\lambda dr \ .
\ee
Therefore the conventional variables correspond to the specific choice of
radial gauge, $N_{\rm S}$, for which the radial variable is precisely
the Schwarzschild $r$ variable. We shall refer to this choice as the
Schwarzschild radial gauge. In the terminology of \cite{ujr:hh} this is an
example of an intrinsic gauge choice. The square of the Schwarzschild gauge
function, $N_{\rm S}{}^2$, coincides with the ``Schwarzschild correction
factor'' of Harrison et al.\cite{htww:grav} This factor determines
whether or not there is a conical singularity at the center of the star
model. It can be computed in an arbitrary radial gauge by the relation (a
comma will denote differentiation throughout the paper)
\be\label{eq:schwarzgauge}
     N_{\rm S} = e^\lambda = N (r_{,R})^{-1}
         = N e^{-\beta^1} (\beta^1_{,R})^{-1} \ .
\ee
As for spacelike slicings the Misner variables\footnote{We use capital indices
($A,B,\ldots$) for the Misner variables taking the values
$0,+$. (The third Misner variable $\beta^-$ is not needed in this context.) }
$\beta^A$ are defined by
\be\eqalign{
        \beta^0 = \case13 (2\beta^1 + \beta^3) \ ,\qquad\qquad
       &\beta^1 = \beta^0 +  \beta^+ \ ,\cr
        \beta^+ = \case13 ( \beta^1 - \beta^3) \ ,
       &\beta^3 = \beta^0 - 2\beta^+ \ .\cr
}\ee
diagonalize the kinetic energy in the ADM Hamiltonian. It is also convenient
to define the Taub radial gauge by $N = N_T = 12e^{3\beta^0}$
\cite{taub:51}.%
\footnote{This gauge was originally introduced by Taub for
spatially homogeneous models. Misner called it the supertime gauge in that
context.}
Using the Lorentz frame $\omega^0 = e^{\beta^3} dt$, $\omega^1 = N_T dR$,
$\omega^2 = e^{\beta^1}\d\theta$,  $\omega^3 = e^{\beta^1}\sin\theta \d\phi$
the Einstein equations for the metric (\ref{eq:metric}) become
\be\eqalign{
      G_{00} &= - \case1{24} e^{-6\beta^0} T
                - \case1{72} e^{-6\beta^0} (\beta^0_{,RR}+\beta^+_{,RR})
                + e^{-2(\beta^0+\beta^+)} = \kappa\rho \ ,\cr
      G_{11} &= - \case1{24} e^{-6\beta^0} T - e^{-2(\beta^0+\beta^+)}
                = \kappa p \ ,\cr
      G_{22} &=   G_{33} = \case1{24} e^{-6\beta^0} T
                + \case1{144} e^{-6\beta^0} (2\beta^0_{,RR} - \beta^+_{,RR})
                = \kappa p \ ,\cr
}\ee
where
\be
        T = \case12 \left[ -(\beta^0_{,R})^2 + (\beta^+_{,R})^2 \right] \ .
\ee
Note that the above equations are given in the Taub radial gauge. Eliminating
$T$ and solving for the second derivatives yields
\be\label{eq:secder}\eqalign{
     \beta^0_{,RR} &=   96 e^{4\beta^0-2\beta^+}
                      + 24\kappa e^{6\beta^0} (5p-\rho) \ ,\cr
     \beta^+_{,RR} &=   48 e^{4\beta^0-2\beta^+}
                      - 48\kappa e^{6\beta^0} (p+\rho) \ .\cr
}\ee

The ADM Hamiltonian for a spherically symmetric static fluid should be
given by \cite{mtw:gravitation,ujr:hh}
\be\label{eq:ADMHam}
    H = -2N \threeV n^\alpha n^\beta
           (G_{\alpha\beta}-\kappa T_{\alpha\beta}) \ ,
\ee
where $\threeV = e^{3\beta^0}$ is the 3-volume element (as usual only
defined only up to a constant factor) and $n^\alpha$ is the unit normal to the
homogeneous hypersurfaces. It is constrained by the Einstein equations to the
zero energy surface, $H=0$. Using the prescription (\ref{eq:ADMHam}) the
Hamiltonian takes the form
\be\label{eq:ham1}
    H = \case12 \Nscr (-p_0{}^2 + p_+{}^2) + 24\Nscr e^{4\beta^0-2\beta^+}
         + 24\kappa \Nscr e^{6\beta^0} \,p(\beta^0,\beta^+) \ ,
\ee
where $\Nscr$ is a relative radial gauge function defined by%
\footnote{
In the context of spacelike slicings Jantzen \cite{jantzen:powerlaw} uses the
notation $x=\Nscr^{-1}$ while Ashtekar \cite{ashtekar:cangrav} uses the term
densitized lapse for $\undersim N = 12\Nscr$.}
$\Nscr = N/N_T$  and $p$ is the pressure written as a function of the metric
variables. As discussed above the Hamiltonian is constrained by the Einstein
equations to the zero energy surface, $H=0$.

At this point it is not yet clear how the variation of the pressure function
$p(\beta^0,\beta^+)$ is to be done in the Hamiltonian framework. In fact it
turns out that the pressure cannot be varied independently with respect to
$\beta^0$ and $\beta^+$. This can be understood if we write down the field
equations obtained from the Hamiltonian (\ref{eq:ham1}) and compare them with
the Einstein equations. For that purpose it is convenient to use the
corresponding Lagrangian in the Taub radial gauge $\Nscr = 1$
\be
    L = \case12 \left[ - (\beta^0_{,R})^2 + (\beta^+_{,R})^2 \right]
         - 24 e^{4\beta^0-2\beta^+}
         - 24\kappa e^{6\beta^0} \,p(\beta^0,\beta^+) \ .
\ee
The equations of motion of this Lagrangian are
\be\label{eq:lageq}\eqalign{
   \frac{\partial L}{\partial\beta^0} -
              \frac{\d}{\d t}\frac{\partial L}{\partial\beta^0_{,R}}
          = - 96 e^{4\beta^0-2\beta^+} - 144\kappa e^{6\beta^0} p
            - 24\kappa e^{6\beta^0} \frac{\partial p}{\partial\beta^0}
            + \beta^0_{,RR} = 0 \ ,\cr
   \frac{\partial L}{\partial\beta^+} -
              \frac{\d}{\d t}\frac{\partial L}{\partial\beta^+_{,R}}
          =   48 e^{4\beta^0-2\beta^+}
            - 24\kappa e^{6\beta^0}\frac{\partial p}{\partial\beta^+}
            - \beta^+_{,RR} = 0 \ .\cr
}\ee
For these equations to coincide with the Einstein equations (\ref{eq:secder})
we see by inspection that the pressure function $p(\beta^0,\beta^+)$ must
satisfy the relations
\be
     \frac{\partial p}{\partial\beta^0} = -(p+\rho) \ ,\qquad
     \frac{\partial p}{\partial\beta^+} = 2(p+\rho) \ ,
\ee
leading to
\be\label{eq:varconstraint}
         2\frac{\partial p}{\partial\beta^0}
        + \frac{\partial p}{\partial\beta^+} = 0 \ .
\ee
This equation shows that $p$ must be considered as a function of $\nu
= \beta^3 =
\beta^0 - 2\beta^+$ in order that Hamilton's equations as derived from the
Hamiltonian (\ref{eq:ham1}) represent the Einstein equations correctly. We
have thus shown that the Hamiltonian for spherically symmetric static models
is given by
\be\label{eq:ham2}
    H = \case12 \Nscr (-p_0{}^2 + p_+{}^2) + 24\Nscr e^{4\beta^0-2\beta^+}
         + 24\kappa \Nscr e^{6\beta^0} \,\Pscr(\nu) \ ,
\ee
where we have introduced the notation $\Pscr(\nu) = p$ for the pressure
function. The momenta can be expressed in terms of the velocities by
\be\label{eq:momenta}\eqalign{
              p_0 = -\Nscr^{-1} \beta^0_{,R} \ ,\cr
              p_+ =  \Nscr^{-1} \beta^+_{,R} \ .\cr
}\ee

A number of remarks about the nature of the Hamiltonian (\ref{eq:ham2}) are
in order. Interestingly, it is the pressure rather than the energy density
which appears in the matter term by contrast to the Hamiltonians of models
with spacelike slicings such as minisuperspace models for example. Since
the kinetic metric carries a Lorentzian signature we can characterize
variables as being timelike, spacelike or null with respect to that metric.
In particular $\beta^1 = \log r$ is a null variable which increases with
distance from the center of the star. It follows that any timelike or null
variable must also be increasing functions. Another useful feature of the
Hamiltonian (\ref{eq:ham2}) is the freedom to perform Lorentz transformations
which preserve the kinetic energy. For some purposes we will
also use null variables defined by
\be\label{eq:null_var}\eqalign{
             w &= \beta^0 + \beta^+ = \beta^1 \ ,\cr
             v &= \beta^0 - \beta^+ = \case13(\beta^1+2\beta^3) \ ,
                  \qquad \nu = \beta^3 = \case12 (-w+3v) \ .\cr
}\ee

The pressure is related to the energy density by the conservation equation
\be\label{eq:cons}
     -\frac{\d p}{\d\nu} = \rho + p \ .
\ee
This is consistent with the variational constraint (\ref{eq:varconstraint}) on
the pressure. Therefore the natural procedure to integrate the system is to
first specify the equation of state $p= p(\rho)$, then use the conservation
equation to write $p$ as a function of $\nu$ and then insert that
function into the Hamiltonian. The resulting Hamilton's equations should then
be integrated with suitable boundary conditions. We shall require as  usual
that the metric is well defined at the center so that it satisfies the
condition of elementary flatness there. In addition it is necessary that the
pressure is zero at the surface of the star in order that the geometry can be
joined smoothly to the Schwarzschild metric in the exterior. To see how to
express the elementary flatness condition analytically in terms of our
variables we write the spatial 3-metric as
\be
   N_{\rm S}{}^2 \d r^2 + r^2 \d\Omega^2 \ .
\ee
Elementary flatness at the center requires that the Schwarzschild radial
gauge function $N_{\rm S}$ (expressed in a general gauge in equation
(\ref{eq:schwarzgauge})) tends to unity \cite{htww:grav}. Using null variables
the condition becomes
\be\label{eq:flatcond}
    N_{\rm S}{}^{-1} = N^{-1} e^w \frac{\d w}{\d R} \rightarrow 1 \ .
\ee

To analyze how the zero pressure limit appears in configuration space let
$\rho_{s}$ be the value of $\rho$ at the surface. It is useful to distinguish
between the cases $\rho_{s}\neq0$ (liquid-like equation of state) and
$\rho_{s}=0$ (gas-like equation of state). In the first case when the
pressure tends to zero it follows from (\ref{eq:cons}) that $\nu$ must
tend to a finite constant, $\nu_{s}$. For a given equation of state and
star mass $m$, let $\Qscr_{\rm s}(m)$ be the point in configuration space
where $p=0$. By varying $m$, the points $\Qscr_{\rm s}(m)$ form a curve in
configuration space representing the star surface for the given equation of
state. Since $p$ is a function of the spacelike variable $\nu$ it follows
that the star surface curve is a timelike plane given by
$\nu=\nu_{s}$. In the case $\rho_{s}=0$ one cannot use
(\ref{eq:cons}) to draw the same conclusions.

The center of the star can be defined as the limit of zero 2-volume,
$e^{2\beta^1} \rightarrow 0$ or equivalently $w = \beta^0+\beta^+ \rightarrow
-\infty$. Thus, this limit is lightlike in configuration space. Just as for
minisuperspace models it is possible to perform a Penrose compactification
scheme for the configuration space (cf.\ \cite{ru:visual}). By the
above remarks it then follows that the star center would be located at
$\scri_-$ or $i_-$. A more refined argument can be made to show that
the center of the star is actually located at $i_-$ but we will not pursue
this point further in this paper.

A global quantity of particular interest is the mass function $m(R)$ given by
\be  m(R) = 4\pi \int_0^R \rho R'^2 dR'
          = 4\pi \int_{-\infty}^{\beta^1} \rho e^{3\beta'^1}\d\beta'^1 \ .
\ee
In particular the total mass of the star is the value of the mass function
at the surface, $M= m(R_{\rm s})$. The total mass can also be computed without
integration by using the matching conditions at the surface for one of the
two metric components which depend on the mass. To see how this works for
the component $e^{2\beta^3} = e^{2\nu}$ we use the notation $Z = e^{2\nu}$
and note that for the Schwarzschild exterior geometry we have $Z = 1-2M/r$.
However, we cannot use this relation as it stands to calculate $M$ because
the value $Z_{\rm s}$ is not invariantly defined since we can always rescale
the time variable by a constant factor. Taking that gauge freedom into
account we have
\be\label{eq:Zschwar}
     Z = B(1-2M/r) \ ,
\ee
where $B$ is some arbitrary constant. To eliminate the gauge factor $B$ we
differentiate (\ref{eq:Zschwar}) with respect to $R$ and then solve for $M$.
Evaluating the resulting expression at the surface yields
\be\label{eq:mass1}
   M = \case12\left[r\left(1+\frac{Z}{r Z_{,r}}\right)^{-1} \right]_{\rm s}
     = \case12\left[r\left(1+\frac{r_{,R}\,Z}{r Z_{,R}}\right)^{-1}
                                                          \right]_{\rm s}\ ,
\ee
for Schwarzschild and arbitrary radial gauges respectively. For the second
way of calculating the mass we make use of the metric coefficient $N_{\rm
S}{}^2 = e^{2\lambda} = (1-2M/r)^{-1}$ the last equality holding for the
external geometry. The value of this coefficient is invariant since the
Schwarzschild radial coordinate itself is invariantly defined. Solving for
the mass gives
\be\label{eq:mass2}
     M = \case12 \left[r(1-N_{\rm S}{}^{-2})\right]_{\rm s} \ .
\ee
Which of the expressions for the mass one uses is a matter of convenience.
We note in passing that the two expressions together imply the identity
\be
    (N_{\rm S}{}^2)_{\rm s}
         = 1 + \left(\frac{r Z_{,R}}{r_{,R}Z}\right)_{\rm s} \ .
\ee

It is sometimes useful to adapt the system to the $\beta^3$ variable. This
is achieved by performing a boost with velocity
$1/2$
\be\eqalign{
     \beta^0 &= \case1{\sqrt3} (2\bar\beta^0 +  \bar\beta^+) \ ,\cr
     \beta^+ &= \case1{\sqrt3} ( \bar\beta^0 + 2\bar\beta^+) \ ,
                        \qquad \beta^3 = \nu = -\sqrt3 \bar\beta^+ \ ,\cr
}\ee
leading to the Hamiltonian
\be\label{eq:hambar}
       H = \case12\Nscr(-\bar p_0{}^2 + \bar p_+{}^2)
             + 24 \Nscr e^{2\sqrt3\bar\beta^0}
             + 24 \Nscr \kappa e^{4\sqrt3\bar\beta^0}
                 e^{2\sqrt3\bar\beta^+} \Pscr(-\sqrt3\bar\beta^+)  \ .
\ee
Note that this mathematical structure is similar to that of a negative
curvature Robertson-Walker model minimally coupled to a scalar field (see
\cite{ujr:hh}). The only difference is in the values of the coefficients of
$\bar\beta^0$. We shall also need the Hamiltonian in the null variables
(\ref{eq:null_var})
\be\label{eq:ham_null1}
      H = - 2\Nscr p_w p_v + 24\Nscr e^{w+3v}
          + 24\kappa\Nscr e^{3(w+v)} \Pscr(\nu) \ ,
\ee
recalling that $\nu$ is given by (\ref{eq:null_var}). When integrating the
solutions it is useful to write the Hamiltonian in terms of the velocities
\be\label{eq:ham_null2}
      H = -\case12 \Nscr^{-1} w_{,R} v_{,R} + 24\Nscr e^{w+3v}
        + 24\kappa\Nscr e^{3(w+v)} \Pscr(\nu) \ .
\ee

Exact solutions can be either symmetry
solutions with the full number of integration constants or submanifold
solutions (also called special or particular solutions) which arises whenever
an invariant submanifold can be expressed as an explicit functional relation
on the phase space. Since the latter solutions are only particular solutions
they cannot in general be made to satisfy the physical boundary conditions.
For the models we are considering in this paper the symmetry solutions are
also general solutions since a single symmetry is always sufficient to
integrate a 2-dimensional Hamiltonian system.

The analysis of the symmetries of finite-dimensional Hamiltonians
constrained to a zero energy surface
\be
     H(q,p) = T(q,p) + U(q) = 0\ ,
\ee
with a quadratic kinetic energy function $T(q,p) = \case12h^{AB}(q) p_A p_B$
is much facilitated by going to the Jacobi gauge $\Nscr = \Nscr_J = |2U|^{-1}$
in which the problem is reduced to finding Killing vectors and Killing
tensors in the Jacobi geometry
\be
     \d s_J{}^2 = 2|U| h_{AB} \d q^A \d q^B \ .
\ee
The method has been described in a series of papers dealing with spatially
homogeneous models (see \cite{urj:geom,ru:kt,ujr:hh} and references therein).

The Jacobi metric for the static spherically symmetric fluid can be read
off from (\ref{eq:ham_null1}) with the result (after rescaling by a numerical
conformal factor)
\be\label{eq:jacobimetric}
    \d s_J{}^2 = -2G \d w \d v \ ,\qquad
           G = e^{w+3v} + \kappa e^{3(w+v)} \Pscr(\nu) \ .
\ee
Any Killing tensor of this metric gives rise to a constant of the motion and
a corresponding Lie symmetry of the Einstein equations. Whether or not such a
symmetry exists depends on the form of the pressure function $\Pscr(\nu)$. In
the next section we will see how such functions can be found.

\section{Killing tensor conditions and pressure functions}
\label{sec:ktcond}
In this section we introduce an ansatz which makes it possible to determine
equations of state for which the Jacobi metric (\ref{eq:jacobimetric}) admits
a second rank Killing tensor. This will include Killing vector cases as well.
Since we are using an ansatz we do not expect to find all possible solutions
but it turns out, however, that we do recover a substantial fraction of the
solutions which have been discussed in the literature for their physical or
mathematical interest. Moreover, we show elsewhere \cite{gr:ministar} that
the ansatz actually gives all possible Killing vector solutions .

The Jacobi metric is (1+1)-dimensional and therefore admits three distinct
types of Killing tensors as shown in \cite{ru:kt}. The Killing tensor
$K_{AB}$ may have either two non-null eigenvectors or a single null
eigenvector. When there are non-null eigenvectors two subcases arise, one
which corresponds to standard Hamilton-Jacobi separability and another one
for which the equations of motion can be decoupled in suitable complex
variables. The possibility of non-Hamilton-Jacobi separation is a
consequence of the indefinite nature of the Jacobi metric.

We now state the necessary and sufficient conditions for the existence of a
second rank Killing tensor in the Jacobi geometry (\ref{eq:jacobimetric}).
A set of null coordinates $(W,V)$ are adapted to a second rank Killing tensor
of the 2-dimensional Lorentzian metric
\be
     ds^2 = -2G(W,V) \d W \d V \ ,
\ee
precisely if one of the following conditions hold \cite{ru:kt}:
\begin{eqnarray}\label{eq:ktcond}
  G_{,WW} = 0  \;\;{\rm or}\;\;  G_{VV} = 0\ ,
            &\hbox{null Killing tensor, linear decoupling,}\nonumber\\
	 G_{,WW}-G_{,VV} = 0 \ ,
            & \hbox{non-null Killing tensor, Hamilton-Jacobi separation,}
                                                           \nonumber\\
	 G_{,WW}+G_{,VV} = 0 \ ,
            & \hbox{non-null Killing tensor, complex decoupling.}
\end{eqnarray}

The Killing vector case is contained in the Hamilton-Jacobi separation case.
As a first step we check if the original null coordinates $(w,v)$ defined in
(\ref{eq:null_var}), modulo a possible scaling, can be made symmetry adapted
in this sense by some appropriate choice of pressure function
$\Pscr(\nu)$. To use the above conditions (\ref{eq:ktcond}) we define null
variables $(W,V)$ by $w=kW$, $v=V$. This corresponds to a boost combined
with a trivial scaling of the original Misner variables $\beta^A$. The metric
function can then be written as
\be
     G = e^{kW+3V} +\kappa e^{3kW+3V} \Pscr(\nu) \ ,
\ee
Calculating the second derivatives and representing them in the variables
$(\nu,V)$, they can be displayed in the form
\be\label{eq:gder_orig}\eqalign{
     e^{-12V+6\nu} G_{,WW} &= k^2 e^{-6V}
       + \kappa k^2 \left[9\Pscr(\nu)-3\Pscr'(\nu)
       +\case14 \Pscr''(\nu)\right]\ ,\cr
     e^{-12V+6\nu} G_{,VV} &= 9 e^{-6V}
       + 9\kappa \left[\Pscr(\nu) + \Pscr'(\nu)
       + \case14 \Pscr''(\nu)\right] \ .\cr
}\ee
It is apparent from these formulas that $G_{,WW}$ and $G_{,VV}$ cannot
separately be set to zero. It follows that there is no null Killing tensor
adapted to these coordinates. Complex decoupling is also impossible since the
coefficient of the first term of the right hand side has the same sign in
both expressions in (\ref{eq:gder_orig}). Consequently that term cannot be
canceled when taking the sum. We can, however, obtain Hamilton-Jacobi
separation by taking the difference
\begin{eqnarray}
   e^{-12V+6\nu}(G_{,WW}-G_{,VV}) = (k^2-9)e^{-6V} \nonumber\\
            + \kappa [9(k^2-1)\Pscr(\nu) - 3(k^2+3)\Pscr'(\nu)
            + \case14(k^2-9)\Pscr''(\nu)] \ .
\end{eqnarray}
It follows that the right hand side vanishes identically precisely when $k^2 =
9$ and $\Pscr'(\nu) = 2\Pscr(\nu)$. Comparison with equation (\ref{eq:cons})
then leads to the equation of state $\rho+3p=0$. However, this is only a
special case of the more general equation of state $\rho+3p = constant$ for
which the general exact solution was found by Whittaker \cite{whittaker:ss}
and which is shown to correspond to a null Killing tensor in section
\ref{sec:null}.

Let us now consider more general transformations than the simple null scaling
used above. To help us in the search for suitable conformal transformations we
examine the structure of the matter potential for some equations of state of
physical interest. For spatially homogeneous cosmological models with the
usual equation of state $p=(\gamma-1)\rho$ the minisuperspace potential is a
sum of terms which are exponentials of linear combinations of the Misner
variables. Integrating the conservation equation (\ref{eq:cons}) one finds
that this is also true for many of the equations of state which have been
discussed in connection with relativistic star models. Most such equations of
state are special cases of the relation
\be\label{eq:eqstategen}
     \rho = A p^{n/(n+1)} + B p \ ,
\ee
which interpolates between a relativistic polytrope $p = K\rho^{1+1/n}$ of
index $n$ for small pressures and a gamma-law equation of state $p =
(\gamma-1)\rho$ for large pressures. The choice $A = K^{-n/(n+1)}$, $B = n$
gives the relativistic ideal gas considered by Tooper \cite{tooper:apj65}.
Other special cases of (\ref{eq:eqstategen}) are $\rho = \rho_0 +
(\gamma-1)^{-1} p$ and an incompressible fluid $\rho = \rho_0$. To this list
can be added Buchdahl's $n=1$ polytrope while Buchdahl's $n=5$ polytrope is
an example of an equation of state which is not of the form
(\ref{eq:eqstategen}). The exponential representations of the above mentioned
equations of state are given in Table \ref{tab:eqstate}. Symmetry adapted
variables for solvable spatially homogeneous models are either linear
combinations of the Misner variables themselves or exponentials of null
Misner variables. Also, as noted in \cite{ujr:hh} the null variables ($W, V$)
defined by $W =e^w$, $V^{3/2} = e^v$ are symmetry adapted for the
Schwarzschild interior solution. In all cases the symmetry adapted potential
turns out to be a polynomial in the symmetry adapted variables. These
observations together lead us to the ansatz
\be\label{eq:ansatz}
       W^\alpha = e^w = r \ ,\quad  V^\beta = e^v\ ,
\ee
for the conformal transformation where $\alpha$ and $\beta$ are constant
parameters to be determined.\footnote{ One could also use the slightly more
general transformation obtained by replacing the first relation by the scaled
form $e^w = (kW)^\alpha$. However, it turns out that this additional freedom
does not lead to any new solvable models. } It will be shown that the above
ansatz for the conformal transformation can be used to recover all (to this
author's knowledge) known equations of state which lead to a general solution
of the equations of motion. In addition the ansatz also leads some new exact
solutions.

\renewcommand{\arraystretch}{1.5}
\begin{table}
\caption{\label{tab:eqstate}\protect\footnotesize
Expressions for pressure functions $p = \Pscr(\nu)$ for some equations of
state where $\rho_0$, $p_0$, $p_*$ and $p_\star$ are constants. The
corresponding fluid potentials are sums of exponentials for integer $n$. The
expressions may be subjected to a gauge translation of the metric variable
$\nu =\beta^3$. The type codes stand for polytrope-gamma law (PG), Tooper's
equation of state (T), relativistic polytrope (RP), incompressible
fluid-gamma law (IG), incompressible fluid (I), gamma law (G) and Buchdahl's
$n=1$ and $n=5$ polytropes (BP1 and BP5). The velocity of sound for
Buchdahl's $n=1$ polytrope is less than the speed of light for pressures $p <
p_*$. Buchdahl's $n=5$ polytrope satisfies $p < \rho$ for $p<p_\star$. }
\vspace{12pt}\footnotesize\rm
\begin{tabular}{@{}lll}
 \hline
    $p = \Pscr(\nu)$ & equation of state & type \\
 \hline
    $(B+1)^{-n-1} A^{n+1} [e^{-(B+1)\nu/(n+1)} - 1]^{n+1}$
                                   & $\rho = A p^{n/(n+1)} + B p$  & PG  \\
    $(n+1)^{-n-1} K^{-n} (e^{-\nu} -1)^{n+1}$
                                    & $\rho = (p/K)^{n/(n+1)}+np$  & T   \\
    $K^{-n} [e^{-\nu/(n+1)} -1]^{n+1}$
                                           & $p = K \rho^{1+1/n}$  & RP  \\
    $(n+1)^{-1} \rho_0 [e^{-(n+1)\nu} - 1]$
                    & $\rho = \rho_0 + np$\ ,  [$n=1/(\gamma-1)$]  & IG  \\
    $\rho_0 (e^{-\nu} - 1)$                       & $\rho=\rho_0$  & I   \\
    $p_0 e^{-(n+1)\nu}$ &$p=(\gamma-1)\rho$\ , [$n=1/(\gamma-1)$]  & G   \\
    $9p_*(e^{2\nu}-1)^2$            & $\rho = 12\sqrt{p_*p} - 5p$  & BP1 \\
    $6^{-6} 7^5 p_\star e^{-\nu}(1-e^{\nu})^6$
              &  $p = \rho^{6/5} /(7p_\star^{1/5} - 6\rho^{1/5})$  & BP5 \\
 \hline
\end{tabular}
\end{table}

In the null variables defined by the ansatz (\ref{eq:ansatz}) the Jacobi
metric (\ref{eq:jacobimetric}) takes the form (upon rescaling by a constant
factor)
\be\label{eq:jacobimetric2}\eqalign{
    ds_J{}^2 &= -2 G \d W \d V \ ,\cr
           G &= W^{\alpha-1}V^{3\beta-1}
              +\kappa W^{3\alpha-1}V^{3\beta-1} f(Z) \ ,\cr
}\ee
and we have defined $Z = e^{2\beta^3} = W^{-\alpha}V^{3\beta}$ and $f(Z) = p
= \Pscr(\frac12\log Z)$. Next we ask under what conditions on the parameters
$\alpha$ and $\beta$ and on the function $f(Z)$ do the variables $W$ and $V$
become symmetry adapted with respect to a Killing vector or a second rank
Killing tensor. For this purpose we use the conditions (\ref{eq:ktcond}) and
we therefore compute the second derivatives of the metric conformal factor
from (\ref{eq:jacobimetric2}) with the result
\begin{eqnarray}\label{eq:gder}
  G_{,WW} =   (\alpha-1)(\alpha-2)W^{\alpha-3}V^{3\beta-1}
        + (3\alpha-1)(3\alpha-2)W^{3\alpha-3}V^{3\beta-1}f(Z) \nonumber\\
        - \alpha(5\alpha-3)W^{2\alpha-3}V^{6\beta-1}f'(Z)
        + \alpha^2 W^{\alpha-3}V^{9\beta-1}f''(Z) \ , \nonumber\\
  G_{,VV} =   (3\beta-1)(3\beta-2)W^{\alpha-1}V^{3\beta-3}
        + (3\beta-1)(3\beta-2)W^{3\alpha-1}V^{3\beta-3}f(Z) \nonumber\\
        + 9\beta(3\beta-1)W^{2\alpha-1}V^{6\beta-3}f'(Z)
        + 9\beta^2 W^{\alpha-1}V^{9\beta-3}f''(Z) \ .
\end{eqnarray}

\section{Null Killing tensor solutions}\label{sec:null}
In this section we determine the equations of state for which there exists a
null Killing tensor in the minisuperspace Jacobi geometry. From the conditions
($i$)--($iii$) in section \ref{sec:ktcond} we know that the Jacobi metric
(\ref{eq:jacobimetric}) admits a null Killing tensor if one of $G_{,WW}$ or
$G_{,VV}$ vanishes identically. Our aim is to find all possible functions $f$
for which this is the case subject to the ansatz (\ref{eq:ansatz}). To
facilitate this analysis we replace the variable set $(W,V)$ in
(\ref{eq:gder}) by $(Z,V)$ with the result
\begin{eqnarray}\label{eq:null_der}
  G_{,WW} =   V^{12\beta-1-9\beta/\alpha}Z^{(3-\alpha)/\alpha}
          \left[ (\alpha-1)(\alpha-2)V^{-6\beta} \right. \nonumber\\
               + \left.(3\alpha-1)(3\alpha-2)Z^{-2}f(Z)
               - \alpha(5\alpha-3)Z^{-1}f'(Z)
               + \alpha^2 f''(Z) \right] \ , \nonumber\\
  G_{,VV} =   V^{12\beta-3-3\beta/\alpha}Z^{(1-\alpha)/\alpha}
          \left[ (3\beta-1)(3\beta-2)V^{-6\beta} \right. \nonumber\\
               + \left. (3\beta-1)(3\beta-2)Z^{-2}f(Z)
               + 9\beta(3\beta-1)Z^{-1}f'(Z)
               + 9\beta^2 f''(Z) \right] \ .
\end{eqnarray}

 From the first of these equations we see by inspection that $G_{,WW} = 0$
precisely if either
\be\label{eq:nc1}
     \alpha=1 \ , \qquad  Z^2 f''(Z) - 2Z f'(Z) + 2 f(Z) = 0 \ ,
\ee
or
\be\label{eq:nc2}
     \alpha=2 \ , \qquad  2 Z^2 f''(Z) - 7 Z f'(Z) + 10 f(Z) = 0 \ .
\ee
In the first of these cases, (\ref{eq:nc1}), the two independent solutions
are $f(Z) = Z$ and $f(Z) = Z^2$ leading to the pressure function
\be\label{eq:np1}
    p = - a e^{4\beta^3} + b e^{2\beta^3} = -aZ^2 + bZ \ ,
\ee
where $a$ and $b$ are integration constants. Using the conservation equation
(\ref{eq:cons}) the energy density is then given by
\be
    \rho = 5a Z^2 - 3b Z \ .
\ee
The equation of state itself can be written in the form
\be\label{eq:nulleqstate1}
    a(5p+\rho)^2 = 2b^2(3p+\rho) \ .
\ee
It can be shown that $p$, $\rho$ and $\d p/ \d\rho$ cannot all be positive for
this equation of state. It is therefore unphysical according to our criterion.
We conclude the examination of this case by noting that the special cases
$a=0$ and $b=0$ correspond to the equations of state $3p+\rho=0$ and
$5p+\rho=0$ respectively. The first of these equations of state was already
encountered above as an exactly solvable case while in the second case the
Jacobi geometry (\ref{eq:jacobimetric2}) is flat.

Moving on to the case (\ref{eq:nc2}), two independent solutions are $f(Z)
= Z^2$ and $f(Z) = Z^{5/2}$. The pressure and energy density can then be
expressed as
\be
     p = -a Z^{5/2} +  b Z^2\ ,\qquad \rho = 6a Z^{5/2} - 5b Z^2 \ .
\ee
The equation of state can be written in the form
\be\label{eq:nulleqstate2}
     a^4 (6p+\rho)^5 = b^5 (5p+\rho)^4 \ .
\ee
Again this an unphysical equation of state for which $p$, $\rho$ and $\d p/
\d\rho$ cannot all be zero simultaneously.

The second possibility for a null Killing tensor occurs if $G_{,VV}=0$ which
happens precisely if either
\be\label{eq:nc3}
     \beta = \case13 \ , \qquad  f''(Z) = 0 \ ,
\ee
or
\be\label{eq:nc4}
     \beta = \case23 \ , \qquad  2 Z f''(Z) + 3 f'(Z) = 0 \ .
\ee
In the first of these cases, (\ref{eq:nc3}), we have $p = -aZ + b$ and $\rho =
3aZ - b$ leading to the equation of state
\be\label{eq:whittaker}
     3p + \rho = 2b \ .
\ee
The corresponding solution was found by Whittaker \cite{whittaker:ss}.
Recall that the special case $b=0$ was shown in section \ref{sec:ktcond} to
admit a non-null Killing tensor. The corresponding Jacobi geometry therefore
admits two inequivalent Killing tensors. Finally in the case given by
(\ref{eq:nc4}) we recover Schwarzschild's interior solution with the
equation of state parametrized as
\be\label{eq:schwar_interior}\eqalign{
              p &= -a + bZ^{-1/2} \ ,\cr
           \rho &= a \ .\cr
}\ee
This concludes the discussion of possible null Killing tensor cases.

\section{Non-null Killing tensor solutions}
To investigate the existence of non-null Killing tensors we must take the sum
and the difference of the second derivatives $G_{,WW}$ and $G_{,VV}$ and set
the resulting expression equal to zero according to the conditions
(\ref{eq:ktcond}). After multiplication by a suitable non-vanishing function
we obtain
\begin{eqnarray}\label{eq:nn_cond}
     Z^{(1-\alpha)/\alpha} V^{12\beta-3-3\beta/\alpha}
               (G_{,WW}-\epsilon G_{,VV})
      = - \epsilon B(Z) V^{\nu_1} + Z^{2/\alpha} A(Z) V^{\nu_2} \nonumber\\
              - \epsilon (3\beta-1)(3\beta-2) V^{\nu_3}
              + (\alpha-1)(\alpha-2) Z^{2/\alpha} V^{\nu_4} = 0\ ,
\end{eqnarray}
where we have introduced the notation
\be\eqalign{
    A(Z) = \alpha^2 f''(Z) - \alpha(5\alpha-3) Z^{-1} f'(Z)
            +  (3\alpha-1)(3\alpha-2) Z^{-2} f(Z) \ ,\cr
    B(Z) = 9\beta^2 f''(Z) + 9\beta(3\beta-1) Z^{-1} f'(Z)
            + (3\beta-1)(3\beta-2) Z^{-2} f(Z) \ ,\cr
    (\nu_1,\nu_2,\nu_3,\nu_4) = (0,\mu,-6\beta,\mu-6\beta) \ ,\cr
}\ee
with $\mu = 2-6\beta/\alpha$, $\epsilon = 1$ for Hamilton-Jacobi
separation and $\epsilon = -1$ for complex decoupling. To analyze when the
above expression vanishes identically we consider the dependence on the
variable $V$. If the exponents $\nu_i$ $(i=1,2,3,4)$ are all different then
we see that the form of the coefficient of $V^{\nu_4}$ in (\ref{eq:nn_cond})
implies that either $\alpha = 1$ or $\alpha = 2$. In the first case we have
$\nu_2 = 2(1-3\beta)$ implying that $\beta \neq 1/3$. Inspection of the
coefficient of $V^{\nu_3}$ then shows that we must have $\beta = 2/3$ implying
\be
    (\nu_1,\nu_2,\nu_3,\nu_4) = (0,-2,-4,-6) \ .
\ee
Equating the remaining two coefficients to zero we obtain
\be\eqalign{
     A(Z) &= f''(Z) - 2 Z^{-1} f'(Z) + 2 Z^{-2} f(Z) = 0 \ ,\cr
     B(Z) &= 4 f''(Z) + 6 Z^{-1} f'(Z) = 0 \ .\cr
}\ee
However, it's easy to see that this pair of equations is incompatible.
Considering now the second case $\alpha=2$, we have
\be
    (\nu_1,\nu_2,\nu_3,\nu_4) = (0,2-3\beta,-6\beta,2-9\beta)) \ .
\ee
Comparison with the coefficient of $V^{\nu_3}$ shows that $\beta=1/3$
leading to
\be
    (\nu_1,\nu_2,\nu_3,\nu_4) = (0,1,-2,-1) \ .
\ee
The conditions for the remaining coefficients then become
\be\eqalign{
     A(Z) &= 4 f''(Z) - 14 Z^{-1} f'(Z) + 20 Z^{-2} f(Z) = 0 \ ,\cr
     B(Z) &=   f''(Z) = 0 \ .\cr
}\ee
This system also turns out to be incompatible. We have thus shown that at
least two of the exponents $\nu_i$ must be equal.

Suppose first that $\nu_1=\nu_2$, i.e., $\alpha=3\beta$. Then
it follows that
\be
    (\nu_1,\nu_2,\nu_3,\nu_4) = (0,0,-6\beta,-6\beta)) \ ,
\ee
implying that there are now only two linearly independent functions of $V$
in (\ref{eq:nn_cond}), namely $V^{\nu_1} = 1$ and $V^{\nu_3} = V^{-6\beta}$.
Setting the coefficient of the latter function equal to zero shows that one
of the following two conditions must hold:
\be\label{eq:nn_cases}\cases{
  \alpha = 1 \qquad \beta = 1/3 \ , & (I)  \cr
  \alpha = 2 \qquad \beta = 2/3 \ . & (II) \cr
}\ee
As we shall see case (I) contains Buchdahl's $n=1$ polytrope solution
\cite{buchdahl:poly_one} as a special case while  case (II) corresponds to
Buchdahl's $n=5$ polytrope solution \cite{buchdahl:poly_five} as recently
generalized by Simon \cite{simon:polyfive}. The corresponding equations of
state will be discussed in the next section.

Going now to the remaining cases we note that $\nu_1 \neq \nu_3$ always
holds since $\beta \neq 0$. Consider therefore $\nu_1 = \nu_4$ i.e.\ $\mu =
6\beta$ leading to
\be
    (\nu_1,\nu_2,\nu_3,\nu_4) = (0,6\beta,-6\beta,0) \ .
\ee
There are now three linearly independent functions of $V$ in
(\ref{eq:nn_cond}). Inspecting the coefficient of one of those functions,
$V^{\nu_3}$, we find that there are two possibilities, $\beta = 1/3$ or
$\beta = 2/3$. The first of these values for $\beta$ is not compatible with
$\mu = 6\beta$. The second value implies $\alpha = -2$. For the remaining
two coefficients to vanish we then have the conditions
\be\eqalign{
   -\epsilon B(Z) + 12 Z^{-1}
         &= -4\epsilon f''(Z) - 6\epsilon Z^{-1} f'(Z) + 12 Z^{-1} = 0\ ,\cr
    A(Z) &= 4 f''(Z) - 26 Z^{-1} f(Z) + 56 Z^{-2} f(Z) = 0 \ .\cr
}\ee
Again, this is an incompatible system.

The next case is $\nu_2 = \nu_3$ corresponding to $\mu =
-6\beta$. We then have
\be
    (\nu_1,\nu_2,\nu_3,\nu_4) = (0,-6\beta,-6\beta,-12\beta) \ .
\ee
There are again three linearly independent functions of $V$ in
(\ref{eq:nn_cond}). In particular, from the coefficient of $V^{\nu_4}$ we
find that either $\alpha=1$ or $\alpha=2$. The first of these values is
incompatible with $\mu = -6\beta$ so we consider now the second. The
remaining equations to be satisfied in that case are
\be\eqalign{
        B(Z) &= 2[2f''(Z) + 9Z^{-1}f'(Z) + 6Z^{-2}f(Z) = 0 \ ,\cr
        Z A(Z) - 12\epsilon
             &= 4 Z f''(Z) - 14 f'(Z)
                      + 20 Z^{-1} f(Z) - 12\epsilon = 0 \ ,\cr
}\ee
Again it can easily be confirmed that this system is incompatible.
Note finally that we always have $\nu_3 \neq \nu_4$. We have therefore
exhausted the possibilities to satisfy (\ref{eq:nn_cond}).

\section{Equations of state corresponding to non-null Killing tensors}
\label{sec:eqns_kt}

We now derive the equations of state corresponding to (I) and (II) of
(\ref{eq:nn_cases}). The equation of state for case (I) will be discussed in
some detail while we shall be content with just writing down the case (II)
equation of state. In case (I) the condition (\ref{eq:nn_cond}) is reduced to
\be
    -\epsilon B(Z) + Z^2 A(Z) = 0 \ ,
\ee
leading to the equation
\be
    (Z^2 - \epsilon) f''(Z) - 2Z f'(Z) + 2 f(Z) = 0 \ .
\ee
Two independent solutions of this equation are $f(Z) = Z$ and $f(Z) = Z^2 +
\epsilon$. It follows that we can formally reobtain the equation state
(\ref{eq:nulleqstate1}) by letting $\epsilon \rightarrow 0$. However, as we
shall see the physical properties of the equation of state is very much
different when $\epsilon$ is nonzero. We write the general solution as
\be
    f(Z) = a Z^2 - 2b Z + a \epsilon \ ,
\ee
corresponding to the pressure function
\be
    p = a e^{4\nu} - 2b e^{2\nu} + a \epsilon \ .
\ee
It follows that pressure and energy density can be parametrized as
\be\label{eq:eqn_st_Z}\eqalign{
         p &=   a Z^2 - 2b Z + a\epsilon \ ,\cr
      \rho &= -5a Z^2 + 6b Z - a\epsilon \ .\cr
}\ee
The equation of state can be written explicitly in the form
\be
    8b^2 (2a\epsilon-3p-\rho) = a(4a\epsilon-5p-\rho)^2 \ .
\ee
Setting $a=0$ we again have the equation of state $3p + \rho =0$. Setting
$b=0$ gives the equation of state $5p+\rho = 4a\epsilon$. It is convenient
to write the equation of state as
\be\label{eq:eqstateB1}\eqalign{
         p &= a(   Z^2 - 2\delta Z + \epsilon) \ ,\cr
      \rho &= a(-5 Z^2 + 6\delta Z - \epsilon) \ ,\cr
}\ee
where $\delta = b/a$. The parameter $a$ is then seen to be a scaling
which does not matter for physical quantities such as $\d p/\d\rho$ or the
relativistic adiabatic index $\gamma = (1+\rho/p)(\d p/\d\rho)$. Its absolute
value is also irrelevant for inequalities such as $p>0$ and energy
conditions such as $\rho > 3p$. However, its sign does matter and must be
determined in each case. For example, suppose the equation of state is given
by $p = a x^2$, $\rho = a x$ corresponding to $p=a^{-1}\rho^2$. For $a>0$ this
is a $n=1$ polytrope while for $a<0$ it is clearly unphysical. Moreover,
given any solution of the Einstein equations $g_{\alpha\beta}$ for any
equation of state $p = \Phi(x)$, $\rho = \Psi(x)$, the scaling properties of
the Einstein equations guarantee that the scaled metric $a^{-1}
g_{\alpha\beta}$ together with the scaled equation of state $p = a\Phi(x)$,
$\rho = a\Psi(x)$ also solve the Einstein equations. In fact, this is the
relativistic version of a mechanical similarity transformation.

For the Hamilton-Jacobi case ($\epsilon =1$) with $\delta =1$ we recover
Buchdahl's $n=1$ polytrope solution \cite{buchdahl:poly_one}. The harmonic
solution and the Hamilton-Jacobi solution with $\delta \neq 1$ have
apparently not appeared in the literature before. We now take a first look
at the physial properties of these equations of state. The velocity of sound
is given by the formula
\be
     v_{\rm sound}^2 = \frac{\d p}{\d\rho} = \frac{\delta-Z}{5Z-3\delta} \ .
\ee
Since $Z>0$ by definition the condition $0 \leq \d p/\d\rho \leq 1$ can
only be satisfied if $\delta > 0$ and
\be\label{eq:Zcond1}
      \frac{2\delta}{3} \leq Z \leq \delta  \ .
\ee
The energy condition
\be
     \rho+p = 4aZ(-Z+\delta) \geq 0 \ ,
\ee
then shows that we must have $a>0$. Considering now the condition $p\geq 0$ we
first note that if $\epsilon =-1$, then that condition is nowhere satisfied
in the region given by (\ref{eq:Zcond1}). Thus the harmonic solution is
nowhere physical according to the adopted criterion and we shall not
consider it further.

For the Hamilton-Jacobi solution, $\epsilon =1$, we find that if $\delta \leq
1$ then $p \geq 0$ automatically while if $\delta >1$ we have the further
condition
\be\label{eq:Zcond2}
     Z \leq \delta - \sqrt{\delta^2-1} \ .
\ee
Compatibility of (\ref{eq:Zcond2}) with (\ref{eq:Zcond1}) now leads to a
restriction on $\delta$ given by $\delta/3 > \sqrt{\delta^2-1}$. Since
$\delta>0$ this is equivalent to
\be
     \delta < \case{3\sqrt2}{4} \approx 1.06 \ .
\ee
Summarizing our analysis so far, the conditions $p\geq0$ and $0\leq \d
p/\d\rho \geq1$ are both satisfied if $\epsilon=1$ and the conditions
\be\label{eq:pv_cond}
     0 < \delta < \case{3\sqrt2}{4} \ , \qquad
     \case{2\delta}{3} \leq Z \leq f(\delta) \ ,
\ee
are satisfied and where $f(\delta)$ is defined by
\be
     f(\delta) = \cases{
      \delta                    &  for $\delta \leq 1$  \ ,\cr
      \delta-\sqrt{\delta^2-1}  &  for $\delta >1$      \ .\cr}
\ee
Examining now the condition $\rho \geq 0$, we find that the polynomial
$-5Z^2 + 6\delta Z - 1$ must have two distinct real roots (excluding the $Z
= constant$ case). It follows that $\delta > \sqrt5/3 \approx 0.745$.
Combining the condition $\rho \geq0$ with (\ref{eq:pv_cond}) gives the final
result that the equation of state is physical according to our criterion in
the range
\be\label{eq:Zrange}
     \frac{2\delta}{3} \leq Z \leq g(\delta) \ ,
\ee
where
\be
     g(\delta) = \cases{(3\delta+\sqrt{9\delta^2-5})/5
                                         & for $\delta \leq 1$  \ ,\cr
                        \delta-\sqrt{\delta^2-1}
                                         & for $\delta >1$ \ .\cr}
\ee
The requirement $2\delta/3 < g(\delta)$ further restricts the parameter values
to the range
\be
     \case34 < \delta < \case{3\sqrt2}{4} \ .
\ee
Furthermore the limit $p \rightarrow 0$ can only be reached if $\delta \geq
1$. Therefore, if $\delta <1$, this equation of state cannot be used all the
way out to the star surface. One way comparing the equation of state with
$\delta \neq1$ with Buchdahl's original equation of state is to compute
the ratio $p_*/\rho_*$ between the pressure and energy density in the limit
$v_{\rm sound} \rightarrow 1$. The result is
\be
    \frac{p_*}{\rho_*} = \frac{9-8\delta^2}{16\delta^2-9} \ .
\ee
This shows that the ratio is smaller for $\delta > 1$ and bigger for $\delta
< 1$. A larger ratio can be interpreted as corresponding to a stiffer
equation of state. Another way of examining the physical nature of the
equation of state is to calculate the relativistic adiabatic index
\be
      \gamma = \frac{p+\rho}p \frac{\d p}{\d\rho} \ .
\ee
It can be shown that $\gamma$ increases with $\delta$. This is also an
indication that equations of state with larger $\delta$ are stiffer or less
compressible.

We turn now to case (II) of (\ref{eq:nn_cases}). The pressure function is
given by
\be
     p = f(Z) = a F_1(Z) + b F_2(Z) \ ,
\ee
where
\be\eqalign{
       F_1(Z) &= Z^2 + \case{10}{3}\epsilon Z + 1 \ ,\cr
       F_2(Z) &= Z^{-1/2} (Z^3 + 15\epsilon Z^2 + 15 Z + \epsilon) \ .\cr
}\ee
Using the variable $Y = Z^{1/2} = e^{\nu}$ the equation of state for the
Hamilton-Jacobi case ($\epsilon = 1$) can be written in parametric form as
(following Simon \cite{simon:polyfive})
\be\label{eq:HJeqstate2}\eqalign{
        p &= (6Y)^{-1} \left[ -\rho_+(1+Y)^6 + \rho_-(1-Y)^6 \right] \ ,\cr
     \rho &= \rho_+ (1+Y)^5 + \rho_-(1-Y)^5 \ ,\cr
}\ee
where
\be\eqalign{
      \rho_+ &= -\case12 (a+6b) \ ,\cr
      \rho_- &= -\case12 (a-6b) \ .\cr
}\ee
The subcase $\rho_+ =0$ is Buchdahl's polytrope of index $n=5$
\cite{buchdahl:poly_five}. The general case $\rho_+ \neq0$ was recently
discussed by Simon \cite{simon:polyfive}.

The equation of state for the harmonic case, $\epsilon =-1$, can be written as
\be\label{eq:harmoniceqstate2}
     p   = (6Y)^{-1} \,{\Re e}[\rho_0(1+iY)^6] \ ,\qquad
    \rho = {\Im m}[\rho_0(1+iY)^5] \ ,
\ee
where $\rho_0$ is a complex constant. We have not found any region where
this equation of state is physical.

\section{Solving the configuration equations}
\label{sec:solve}

In this section we outline the procedure for integrating the Hamiltonian
configuration equations for equations of state which correspond to a
Jacobi geometry which admits a second rank Killing tensor. We begin by
expressing the Hamiltonian (\ref{eq:ham_null2}) in terms of the symmetry
adapted variables defined by the ansatz (\ref{eq:ansatz})
\be\label{eq:ham_ad_null}
    H = -\case12 W_{,R} V_{,R} + 24 \alpha\beta
        \left[ W^{\alpha-1}V^{3\beta-1}
                + \kappa W^{3\alpha-1} V^{3\beta-1} f(Z) \right] \ ,
\ee
where the relative radial gauge is given by
\be
     \Nscr = \alpha \beta (WV)^{-1}
           = 12\alpha\beta W^{-(\alpha+3\beta)/(3\beta)} Z^{-1/(3\beta)}\ ,
\ee
while the absolute radial gauge becomes
\be\label{eq:lapse_ad_null}
    N = N_T \Nscr = 12\alpha\beta W^{(3\alpha-2)/2} V^{(3\beta-2)/2} \ .
\ee
We are interested in how the elementary flatness condition
(\ref{eq:flatcond}) restricts the integration constants as we perform the
integration. For that purpose we need to express the Schwarzschild radial
gauge (\ref{eq:schwarzgauge}) in the symmetry adapted variables
\be\label{eq:schwarzgauge2}\eqalign{
     N_{\rm S}{}^{-1}
        &= \alpha N^{-1} W^{\alpha-1} W_{,R}
               = \case1{12}\beta^{-1} W^{-\alpha/2}V^{(2-3\beta)/2}W_{,R} \cr
        &= \case1{12}\beta^{-1} W^{\alpha(1-3\beta)/(3\beta)}
                                            Z^{(2-3\beta)/(6\beta)}W_{,R} \ .
}\ee

In the symmetry adapted variables the Hamiltonian (\ref{eq:ham_ad_null}) has
the form
\be\label{eq:HlinearV}
    H = -\case12 W_{,R}V_{,R} + A(W)V + B(W) \ .
\ee
The procedure to obtain the general solution for this type of Hamiltonian has
been outlined in \cite{ujr:hh}. The fact that the potential in
(\ref{eq:HlinearV}) is linear in one of the null variables, ($V$), leads to a
decoupled equation of motion for the other null variable, ($W$),
\be
     W_{,RR} = 2A(W) \ .
\ee
This equation has the first integral
\be
     E = \case12 \left(W_{,R}{}^2\right) -2\int A(W)\d W \ .
\ee
By integrating this equation we can  express $W$ as a function of $R$.
Inserting this into the Hamiltonian constraint results in the linear first
order equation for $V$ given by
\be
     V_{,R} = (2/W_{,R}) [A(W)V + B(W)] \ .
\ee
The solution of this equation is
\be\label{eq:Vint}
     V = 2 W_{,R} \int B \left(W_{,R}\right)^{-2} \d R \ .
\ee

Rather than deriving the metrics for all the possible models we shall give
a few examples which illustrate the technique. First of all we show how to
calculate the metrics for Schwarzschild's exterior and interior solutions in
the Hamiltonian framework. After that we calculate the metric for the
generalization of Buchdahl's $n=1$ polytrope solution. In the final
subsection we discuss how the remaining models can be integrated. The
equations of state which correspond to minisuperspace Killing tensors are
summarized in table \ref{tab:ss}.

\renewcommand{\arraystretch}{1.9}
\begin{table}
\caption{\label{tab:ss}\protect\footnotesize
Perfect fluids corresponding to Killing tensor solutions. In the first two
columns a ``-" means that the value of $\alpha$ or $\beta$ is arbitrary. The
parameters $a$, $b$, $\rho_\pm$ are real constants while $\rho_0$ is a
complex constant. The notations $Y=e^\nu$ and $Y_\pm=1\pm Y$ are also used.
For the case $\alpha=2$, $\beta=2/3$, it is not possible to write
down an equation of state in closed form. Instead the expressions for
$\rho(Y)$ are given in the fourth column. The notation ``H-J" in the last
column stands for a Hamilton-Jacobi Killing tensor type. The Hamilton-Jacobi
case with $(\alpha,\beta)=(1,1/3)$ reduces to a Killing vector solution when
$a=b$. }
\vspace{12pt}\footnotesize\rm
\begin{tabular}{@{}lllll}
 \hline
     $\alpha$ & $\beta$ & $p=p(Y)$ & \begin{tabular}[l]{l}
                                          equation of state \\
                               {[}$\rho = \rho(Y)$ in the last two rows{]} \\
                                         \end{tabular}
                                       & Killing tensor type \\
 \hline
	    1 & - & $-aY^4+bY^2$ & $a(5p+\rho)^2 = 2b^2(3p+\rho)$ & null \\
     2 & - & $-aY^5+bY^4$ & $a^4(6p+\rho)^5 = b^5(5p+\rho)^4$ & null \\
     - & $1/3$ & $-aY^2+b$ & $3p+\rho=2b$ & null \\
     - & $2/3$ & $-a+bY^{-1}$ & $\rho=a$ & null \\ \ms
	    1 & $1/3$ & $aY^4-2bY^2+a$
               & $\displaystyle\frac{(5p+\rho-4 a)^2}{2 a-3p-\rho}
                                 = \frac{8b^2}{a}$ & H-J \\ \bs
	    1 & $1/3$ & $aY^4-2bY^2-a$
               & $\displaystyle\frac{(5p+\rho+4a)^2}{2a+3p+\rho}
                                 =-\frac{8b^2}{a}$ & harmonic \\ \ms
     2 & $2/3$ & $(6Y)^{-1}\left(-\rho_+Y_+^6+\rho_-Y_-^6\right)$
               & $\rho_+Y_+^5+\rho_-Y_-^5$  &  H-J \\
     2 & $2/3$ & $(6Y)^{-1} \,{\Re e}\left[\rho_0(1+iY)^6\right]$
               & ${\Im m}[\rho_0(1+iY)^5]$  & harmonic \\
 \hline
\end{tabular}
\end{table}

\subsection{Schwarzschild's exterior solution}
To illustrate how the Hamiltonian method works for the static models we first
derive the Schwarzschild vacuum solution. In this case there is no matter term
and choosing $\alpha=1$, $\beta=1/3$, the Hamiltonian (\ref{eq:ham_ad_null})
becomes
\be\label{eq:ham_schwarz}
     H = -\case12 W_{,R} V_{,R} + 8 \ .
\ee
The problem is now trivially solvable. In fact the Jacobi geometry is
obviously flat in this case. The Hamiltonian constraint gives $W_{,R} = C$ and
$V_{,R} = 16/C$ where $C$ is a separation constant. Since $r=e^{\beta^1} = W$
it follows that $R = C^{-1} r$ (up to a gauge translation) and $V = 16 C^{-1}
(R - R_0)$. The radial gauge is given by $N = N_T \Nscr = 4 W^{1/2} V^{-1/2}$.
Using the relations to the metric variables, $e^{2\beta^3} = W^{-1} V$, $N^2
= 16 W V^{-1}$, one then arrives at the usual form of the Schwarzschild
metric with the mass given by $M = CR_0/2$.

\subsection{Schwarzschild's interior solution, $\rho = \rho_{\rm s}$}
The equation of state is parametrized according to (\ref{eq:schwar_interior})
as $p = f(Z) = -a + bZ^{-1/2}$ and $\rho = \rho_{\rm s} = a$. Further we
have $\beta = 2/3$ and we set $\alpha=1$ implying $W=r$ and $Z = W^{-1} V^2$.
Referring to (\ref{eq:HlinearV}) we also have
\be\label{eq:AB}
    A(W) = 16 (1 - \kappa a W^2) \ ,\qquad
    B(W) = 16 \kappa b W^{5/2} \ .
\ee
Applying the procedure outlined above we have the following first integral
\be\label{eq:schwar_integral}
    E = \case12 W_{,R}{}^2 - 32(W - \case13 \kappa a W^3) \ .
\ee
The next step is to use the condition of elementary flatness
(\ref{eq:flatcond}) to determine the value of $E$. It follows form
(\ref{eq:schwarzgauge2}) that $W_{,R}$ can be expressed in terms of the
Schwarzschild radial gauge as
\be\label{eq:schwarzWR}
    W_{,R} = 8 W^{1/2} N_{\rm S}{}^{-1} \ .
\ee
Elementary flatness requires that $N_{\rm S} \rightarrow 1$ as $r = W
\rightarrow 0$. This implies that $W_{,R} =0$ at the center of the star.
It then follows that $E=0$ for regular models. The equation
(\ref{eq:schwar_integral}) can then be integrated further in terms of
elementary functions. Alternatively one can use an intrinsic variable
\cite{ujr:hh} such as $W$ or a function of $W$ as coordinate in place of $R$.
In fact, it is the intrinsice variable choice $W=r$ which gives us the usual
form of the metric. In practice, what we must do to use an intrinsic variable
is to change the integration variable in (\ref{eq:Vint}). Using $W$ as the new
integration variable that equation reads
\be
    V = 2 W_{,R} \int B(W) \left(W_{,R}\right)^{-3} \d W \ .
\ee
Performing the integration gives
\be
     V = \case12a W^{1/2} \left(3b^{-1}
                  - \kappa K\sqrt{1-\case13\kappa a W^2} \right) \ ,
\ee
where $K$ is an integration constant. Using $Z= W^{-1}V^2$ gives one of the
metric coefficients
\be\label{eq:schwarzZ}
     Z = e^{2\nu} = \case14b^2 \kappa K \left(3D
                  - \sqrt{1-\case13\kappa \rho_{\rm s} r^2} \right)^2 \ ,
\ee
where $D = (\kappa a K)^{-1}$. Finally, using (\ref{eq:schwarzWR}) to
compute the remaining metric coefficient yields
\be\label{eq:schwarzlambda}
     e^{2\lambda} = N_{\rm S}{}^2 = 64 W \left(W_{,R}\right)^{-2}
                  = \left(1 - \case13\kappa \rho_{\rm s} r^2\right)^{-1} \ .
\ee
Using (\ref{eq:mass2}) we also obtain the mass from this expression as
$M = \kappa \rho_{\rm s} r_{\rm s}{}^3$. The physical meaning of the
constant $D$ comes from evaluating (\ref{eq:schwarzZ}) at the surface. This
gives $D = (1-\frac13 \kappa \rho_{\rm s} r_{\rm s}{}^2)^{1/2}$ where we have
used $Z_{\rm s} = (b/a)^2$. These relations together with (\ref{eq:schwarzZ})
and (\ref{eq:schwarzlambda}) give the standard form of Schwarzschild's
interior metric as given in \cite{ksmh:exact}. The constant $b$ corresponds
to a gauge scaling of the time coordinate and consequently has no physical
meaning.

\subsection{Buchdahl's polytrope of index one and its generalization}
\label{sec:poly_one}
In this section we calculate the metric components for the solution
corresponding to case (I) of (\ref{eq:nn_cases}). This case generalizes
Buchdahl's $n=1$ polytropic solution to a fluid which is no longer gas-like
in the limit $p=0$ but rather has a non-zero density there, $\rho_{\rm s}
\neq 0$. Since $\alpha = 1$ and $\beta = 1/3$ we have $Z = V/W$ and the
symmetry adapted null variables are $W = e^w$ and $V = e^{3v}$. Specializing
the Hamiltonian (\ref{eq:ham_ad_null}) to this case yields
\be
    H = -\case12 W_{,R} V_{,R}
        + 8 \left[ 1 + \kappa a( V^2 - 2\delta W V + W^2) \right] \ .
\ee
Separation is achieved by introducing the symmetry adapted non-null
variables $(T,X)$ by
\be\label{eq:TX_def}\eqalign{
     T &= \case12 (W + V) = \case12 e^{\beta^1} (1+e^{2\beta^3})
                          = \case12 r(1+Z) \ ,\cr
     X &= \case12 (W - V) = \case12 e^{\beta^1} (1-e^{2\beta^3})
                          = \case12 r(1-Z) \ .\cr
}\ee
In terms of these variables the Hamiltonian takes the manifestly separable
form
\be\label{eq:hamsepB1}
    H = \case12 \left( -T_{,R}{}^2 + X_{,R}{}^2 \right) + 8 \left[
         1 - 2\kappa a(\delta-1) T^2 + 2\kappa a(\delta+1) X^2 \right] \ ,
\ee
leading to the separated equations
\be\eqalign{
     T_{,R}{}^2 + 32\kappa a(\delta-1) T^2 &= K + 16 \ ,\cr
     X_{,R}{}^2 + 32\kappa a(\delta+1) X^2 &= K \ ,\cr
}\ee
where $K$ is the separation constant. It is obvious from the second of these
equations that $K>0$. It is also clear that the solutions $T(R)$ will have
different functional forms for the three cases $\delta < 1$, $\delta = 1$
and $\delta > 1$. Since $T$ is a cyclic variable in the Hamiltonian
(\ref{eq:hamsepB1}) when $\delta = 1$ it follows that the original Buchdahl
$n=1$ polytrope solution corresponds to the existence of a Killing vector in
the minisuperspace Jacobi geometry.

As explained in section \ref{sec:eqns_kt}, when $\delta < 1$ the equation of
state has no zero pressure limit. Although such an equation of state could
still be useful for high pressures we restrict the following discussion to the
cases with $\delta \geq 1$. The solutions are given by
\be\label{eq:TXsolutions}\eqalign{
     T(R) &= \cases{
     4 (\cosh\zeta) (R-R_-)                \ , & for $\delta=1$ \ ,\cr
     4 (\cosh\zeta) \omega_-{}^{-1}\sin[\omega_-(R-R_-)]
                                           \ , & for $\delta>1$ \ ,\cr} \cr
     X(R) &= 4 (\sinh\zeta) \omega_+{}^{-1}\sin[\omega_+(R-R_+)] \ ,    \cr
}\ee
where $\zeta = \arcsinh(\sqrt{K}/4)$, $\omega_\pm
=\sqrt{32a\kappa(\delta\pm1)}$, and the parameters $R_\pm$ are integration
constants. Let us now consider the possible range of values which $T$ and $X$
can assume. It is obvious from (\ref{eq:TX_def}) that $T\geq0$. Also, it
follows from (\ref{eq:Zrange}) that $Z\leq1$ throughout the interior of the
star implying that $X\geq0$. It is also clear from (\ref{eq:TX_def}) that both
$T$ and $X$ vanish at the center. Next we use the translational freedom in the
radial coordinate to make it vanish at the center, $R_{\rm c}=0$. The
condition $T(0) =0$ then implies $R_-=0$ for $\delta =1$ and that
$\omega_-R_-$ is a multiple of $\pi$ when $\delta>1$. Further $X(0) =0$
implies that $\omega_+R_+$ must be a multiple of $\pi$. Since
$\alpha\beta = 1/3 > 0$ it is clear from (\ref{eq:lapse_ad_null}) that the
radial gauge function is positive and hence $R$ increases outward from the
center. To ensure the positivity of $T$ and $X$ when $R>0$ it follows that
$\omega_+ R_+$ and $\omega_- R_-$ must both be even multiples of $\pi$. But
then we can just as well set $R_\pm=0$ without loss of generality.

We now consider the elementary flatness condition (\ref{eq:flatcond}) at the
center of the stellar model. In this case the condition can be transformed to
\be
    \lim_{R\rightarrow 0}  \case14 W_{,R}\, Z^{1/2} = 1 \ .
\ee
For the above solution we have $(T_{,R})_{\rm c} = 4\cosh\zeta$,
$(X_{,R})_{\rm c} = 4\sinh\zeta$ leading to $(W_{,R})_{\rm c} = 4e^{\zeta}$
and $(V_{,R})_{\rm c} = 4e^{-\zeta}$. The central value of $Z$ is given by the
indeterminate expression $Z_{\rm c} = (V/W)_{\rm c}$ which by the previous
argument is seen to be given by $Z_{\rm c} = e^{-2\zeta}$. Using these central
values we find that the elementary flatness condition is already satisfied
without restrictions on the remaining integration constant $\zeta$.
Restricting attention now to the case $\delta >1$, the final form of the
solution is therefore given by
\be\label{eq:buchdahlTX}\eqalign{
     T(R) &= 4 (\cosh\zeta) (\omega_-)^{-1} \sin(\omega_- R) \ ,\cr
     X(R) &= 4 (\sinh\zeta) (\omega_+)^{-1} \sin(\omega_+ R) \ .\cr
}\ee

Referring to (\ref{eq:Zcond1}) and (\ref{eq:Zcond2}) we must require for
consistency that $2\delta/3 < Z_{\rm c} < Z_{\rm s}$ leading to a restriction
on the separation parameter
\be
      2\delta/3 < e^{-2\zeta} < \delta - \sqrt{\delta^2-1} \ .
\ee
The central values of the pressure and energy density can be inferred from
(\ref{eq:eqstateB1}) as functions of $\delta$ and $\zeta$. Using
(\ref{eq:lapse_ad_null}) to calculate the radial gauge we find that the
metric for this model is given by
\be
     \d s^2 = - Z \d t^2 + 16 Z^{-1} \d R^2 + W^2 \d\Omega^2 \ ,
\ee
where $Z = V/W = (T-X)/(T+X)$, $W = T+X$ and $T$ and $X$ are given by
(\ref{eq:buchdahlTX}). The physical properties of this model will be given
elsewhere.

\subsection{Other models}
We begin this subsection by discussing how the remaining null Killing tensor
models can be integrated. Whittaker's solution with equation of state given by
(\ref{eq:whittaker}) has a structure which is similar to Schwarzschild's
interior solution. The Hamiltonian is again linear in $V$ and thus the
solution can be integrated in an analogous manner. In this case the condition
for elementary flatness does not restrict the integration constant $E$ to be
zero but the resulting equations can nevertheless be integrated in terms of
elementary functions in the Schwarzschild radial gauge. This leads back to
the form for the solution given by Whittaker \cite{whittaker:ss}. For the
null Killing tensor cases corresponding to the equations of state
(\ref{eq:nulleqstate1}) and (\ref{eq:nulleqstate2}) the roles of $W$ and $V$
are reversed. This gives a Hamiltonian which is linear in $W$ leading to a
decoupled equation in $V$. Although the integration proceeds in an analogous
way it is less straightforward to impose the elementary flatness condition.
This is because that condition is expressed in terms $W$. However, an
argument can be made using the Hamiltonian constraint to express the flatness
condition in terms of $V$. These solutions cannot in general be expressed in
Schwarzschild coordinates.

The second Hamilton-Jacobi solution corresponding to the equation of state
(\ref{eq:HJeqstate2}) can be integrated in much the same way as Buchdahl's
$n=1$ polytrope and its generalization were treated in section
\ref{sec:poly_one}. Buchdahl's equation of state corresponding to
$\rho_+ =0$ tends to a Newtonian $n=5$ polytrope in the zero pressure limit.
In fact, like the Newtonian $n=5$ polytrope, the star has an inifinite radius.
However, the more general Simon solutions do not seem to have this property
and a closer examination of the physical properties of this class of
solutions would be of interest.

Although the harmonic solutions found in this work do not seem to of
physical interest we briefly indicate the method by which they can be
integrated. The decoupling variables in these cases are complex and given by
$Q = W + iV$ and its complex conjugate $\bar Q = W - iV$. Using these
variables together with the pressure functions (\ref{eq:eqstateB1}) (with
$\epsilon =1$) and (\ref{eq:harmoniceqstate2}) the Hamiltonian
(\ref{eq:ham_ad_null}) becomes explicitly separated into a $Q$-dependent and
a $\bar Q$-dependent part respectively. The procedure to obtain the metric
variables then parallels that for the Hamilton-Jacobi case.

\section{Concluding remarks}
We have shown that a number of known exact solutions for static star
configurations can be described in a unified setting using an ADM-like
minisuperspace geometrization of the dynamics. New solutions have also been
found in this process illustrating the power of the Hamiltonian formalism.
The remark by Kramer et al.\ in 1980 (\cite{ksmh:exact}, p.130-131) that ``A
Hamiltonian formulation \ldots is not well-adapted to searching for exact
solutions" reflected a prejudice of the time which should be dispelled by
now. Indeed, the traditional view towards exact solutions is that, as stated
by Schutz in 1985 when discussing Buchdahl's $n=1$ polytropic
solution, that ``Finding such [exact] solutions is an art which requires the
successful combination of useful coordinates, simple geometry, good
intuition, and in most cases luck" (\cite{schutz:gr}, p.263). In the light of
the present work as well as earlier work on spatially homogeneous models (see
\cite{ujr:hh} and references therein), the opposite view seems perhaps more
appropriate. At least those exact solutions which correspond to symmetries
and also many submanifold solutions \cite{jur:hhscalar} can certainly be
derived by systematic methods.


\begin{thebibliography}{10}

\bibitem{schwarz:interior}
K. Schwarzschild, Sitz.\ Preuss.\ Akad.\ Wiss.  424  (1916).

\bibitem{lm:indexlimit}
L. Lindblom and A.~K.~M. {Masood-ul-{A}lam},  in {\em Directions in General
  Relativity II: Papers in Honor of {D}ieter {B}rill}, edited by B.~L. Hu and
  T.~A. Jacobson (Cambridge University Press, Cambridge, 1993), p.\ 172.

\bibitem{simon:polyfive}
W. Simon, \grg {\bf 26}, 97 (1994).

\bibitem{urj:geom}
C. Uggla, K. Rosquist, and R.~T. Jantzen, \prd {\bf 42},  404  (1990).

\bibitem{ru:kt}
K. Rosquist and C. Uggla, \jmp {\bf 32},  3412  (1991).

\bibitem{ujrz:late}
C. Uggla, R.~T. Jantzen, K. Rosquist, and H. von Zur-M{\umlaut u}hlen, \grg
  {\bf 23},  947  (1991).

\bibitem{atu:miniquant}
A. Ashtekar, R. Tate, and C. Uggla, \ijmpd {\bf 2},  15  (1993).

\bibitem{ru:visual}
K. Rosquist and C. Uggla, \mpla {\bf 8},  2815  (1993).

\bibitem{mtw:gravitation}
C.~W. Misner, K.~S. Thorne, and J.~A. Wheeler, {\em Gravitation} (Freeman, San
  Fransisco, USA, 1973).

\bibitem{ruj:vac}
K. Rosquist, C. Uggla, and R.~T. Jantzen, \cqg {\bf 7},  611  (1990).

\bibitem{ruj:mat}
K. Rosquist, C. Uggla, and R.~T. Jantzen, \cqg {\bf 7},  625  (1990).

\bibitem{tolman:statsol}
R.~C. Tolman, \pr {\bf 55},  364  (1939).

\bibitem{knutsen:gasspheres}
H. Knutsen, \grg {\bf 22},  925  (1990).

\bibitem{ksmh:exact}
D. Kramer, H. Stephani, M.~A.~H. Mac{C}allum, and E. Herlt, {\em Exact
  Solutions of the {E}instein Equations} (VEB Deutscher Verlag der
  Wissenschaften, Berlin, GDR, 1980).

\bibitem{rosquist:kt_sym}
K. Rosquist, \jmp {\bf 30},  2319  (1989).

\bibitem{buchdahl:poly_one}
H.~A. Buchdahl, \apj {\bf 147},  310  (1967).

\bibitem{ujr:hh}
C. Uggla, R.~T. Jantzen, and K. Rosquist, in preparation (unpublished).

\bibitem{htww:grav}
B.~K. Harrison, K.~S. Thorne, M. Wakano, and J.~A. Wheeler, {\em Gravitation
  theory and gravitational collapse} (University of Chicago Press, Chicago,
  USA, 1965).

\bibitem{taub:51}
A.~H. Taub, \am {\bf 53},  472  (1951).

\bibitem{jantzen:powerlaw}
R.~T. Jantzen, \prd {\bf 37},  3472  (1988).

\bibitem{ashtekar:cangrav}
A. Ashtekar, {\em Lectures on non-perturbative canonical gravity} (World
  Scientific, Singapore, 1991).

\bibitem{gr:ministar}
M. Goliath and K. Rosquist, Report, Department of Physics, Stockholm
  University,  (unpublished), in preparation.

\bibitem{whittaker:ss}
J.~M. Whittaker, \prsla {\bf 306},  1  (1968).

\bibitem{tooper:apj65}
R.~F. Tooper, \apj {\bf 142},  1541  (1965).

\bibitem{buchdahl:poly_five}
H.~A. Buchdahl, \apj {\bf 140},  1512  (1964).

\bibitem{schutz:gr}
B.~F. Schutz, {\em A first course in general relativity} (Cambridge University
  Press, Cambridge, U.K., 1985).

\bibitem{jur:hhscalar}
R.~T. Jantzen, C. Uggla, and K. Rosquist, \grg {\bf 25},  409  (1993).

\end{thebibliography}
\end{document}